\documentclass[a4paper]{aa}
\usepackage{graphicx,float}

\def \src {X\thinspace1624$-$490}
\def \degmark{^\circ}
\def \nh {N${\rm _H}$}
\def \ergsec{\hbox{erg s$^{-1}$}}
\def \hcm {\hbox {\ifmmode $ atom cm$^{-2}\else atom cm$^{-2}$\fi}}
\def \arcmin {\hbox{$^\prime$}}
\def \arcsec {\hbox{$^{\prime\prime}$}}

\def\approxgt{\mathrel{\hbox{\rlap{\lower.55ex \hbox {$\sim$}}
        \kern-.3em \raise.4ex \hbox{$>$}}}}
\def\approxlt{\mathrel{\hbox{\rlap{\lower.55ex \hbox {$\sim$}}
        \kern-.3em \raise.4ex \hbox{$<$}}}}
\newcommand{\mc}{\multicolumn}

\begin{document}

\title{Discovery of narrow X-ray absorption features 
from the dipping low-mass X-ray binary \src\ with XMM-Newton}

\author{A. N. Parmar
        \and T. Oosterbroek
        \and L. Boirin
        \and D. Lumb
}
\offprints{A.N. Parmar, \email{aparmar@rssd.esa.int}}

\institute{
       Astrophysics Division, Research and Scientific Support 
       Department of ESA, ESTEC,
       Postbus 299, NL-2200 AG Noordwijk, The Netherlands
}
\date{Received: 12 November 2001 / Accepted: 22 February 2002}

\authorrunning{A.N. Parmar et al.}

\titlerunning{XMM-Newton observations of \src}

\abstract{We report the discovery of narrow X-ray absorption
features from the dipping low-mass X-ray binary \src\ during an
XMM-Newton observation in 2001 February. The features are identified
with the K$\alpha$ absorption lines of Fe\,{\sc xxv} 
and Fe\,{\sc xxvi} and have
energies of $6.72 \pm 0.03$~keV and $7.00 \pm 0.02$~keV 
and equivalent widths (EWs) of 
$-7.5 \, ^{+1.7} _{-6.3}$~eV and $-16.6 \, ^{+1.9} _{-5.9}$~eV, respectively.
The EWs show no obvious dependence on orbital phase, except
during a dip, 
and correspond to a column of
$\approxgt$$10^{17.3}$~Fe~atom~cm$^{-2}$.
In addition,
faint absorption features tentatively identified with 
Ni\,{\sc xxvii} K$\alpha$ and Fe\,{\sc xxvi} K$\beta$ may be present. 
A broad emission feature at $6.58 \, ^{+0.07} _{-0.04}$~keV with an 
EW of $78 \, ^{+19} _{-6}$~eV
is also evident. This is probably
the 6.4~keV feature reported by earlier missions since fitting
a single Gaussian to the entire Fe-K region gives an energy
of $6.39 \, ^{+0.03} _{-0.04}$~keV. 
A deep absorption feature is present during the dip with an energy
consistent with Fe\,{\sc xxv} K$\alpha$.
This is the second dipping LMXRB source from
which narrow Fe absorption features have been observed.  Until recently
the only X-ray binaries known to exhibit narrow X-ray absorption lines
were two superluminal jet sources and it had been suggested that these
features are related to the jet formation mechanism. 
It now appears likely that ionized absorption features
may be common characteristics of accreting systems with accretion disks. 
  } \maketitle

\section{Introduction}
\label{sect:intro}

The low-mass X-ray binary (LMXRB) source \src\ exhibits irregular dips in 
X-ray intensity that repeat every orbital cycle of 
21~hr (Jones \& Watson~\cite{j:85}). 
This long orbital period means that as well as being the most
luminous dip source, \src\ also has the largest stellar
separation and is often referred to as the ``Big Dipper''. 
It is generally accepted that these dips are due to obscuration in 
the thickened outer regions of an accretion disk (White \& Swank~\cite{ws:82}).
The depth, duration and spectral properties of the dips 
vary from source to source and from cycle to cycle (see 
Parmar \& White~\cite{p:88} and  White et al.~\cite{w:95} for reviews).
The \src\ X-ray lightcurve exhibits prominent flares, but no X-ray bursts
have been observed (e.g., Ba\l uci\'nska-Church et al.~\cite{bc:01}, 
Smale et al.~\cite{s:01}).
The X-ray continuum of \src\ may be modeled using
blackbody and Comptonized components which are
absorbed by a column of $\sim$$9 \, 10^{22}$~atom~cm$^{-2}$
(Ba\l uci\'nska-Church et al.~\cite{bc:00}). There is evidence for
the presence of a dust scattering halo (Angelini et al.~\cite{a:96}).
During dips the point-like blackbody is rapidly obscured and the 
extended Comptonized component, presumably an
accretion disk corona, is progressively covered by an extended absorber
(Ba\l uci\'nska-Church et al.~\cite{bc:00}).

XMM-Newton is proving to be very successful in improving our 
understanding of the nature of emitting and absorbing regions in
LMXRB. Reflection Grating Spectrometer  
(RGS) observations of the eclipsing and dipping system EXO\thinspace0748-676,  
revealed the presence of a rich variety
of emission lines identified with O\,{\sc viii} Ly$\alpha$, O\,{\sc vii} 
He-like complex, Ne\,{\sc x}~Ly$\alpha$, Ne\,{\sc ix} He-like complex 
and N\,{\sc vii} Ly$\alpha$ (Cottam et al.~\cite{c:01}).  
In addition, photo-electric absorption 
edges of both O\,{\sc viii} and O\,{\sc vii} were detected
as well as narrow radiative recombination continua of O\,{\sc viii}
and O\,{\sc vii} near their respective absorption edges.  
The line features are significantly broadened
with velocity widths of 1000--3000~km~s$^{-1}$, depending on the
line, and no obvious velocity shift ($<$300~km~s$^{-1}$).
There is no obvious dependence of the line 
properties on orbital phase (even during X-ray eclipse).
European Photon Imaging Camera
(EPIC) and RGS spectra of the eclipsing and dipping system
MXB\thinspace1658$-$298 have revealed the presence of narrow 
resonant absorption features identified with 
O\,{\sc viii}~K$\alpha$, 
K$\beta$, and K$\gamma$, Ne\,{\sc x} K$\alpha$, Fe\,{\sc xxv}~K$\alpha$,
and Fe\,{\sc xxvi}~K$\alpha$,
together with a broad Fe emission feature at $6.47 \, ^{+0.18} _{-0.14}$~keV
(Sidoli et al.~\cite{st:01}).
Again, the properties of these features show no obvious 
dependence on orbital phase, even during dipping intervals. 

Until recently, the only X-ray binaries known to exhibit such narrow 
X-ray absorption 
lines were two superluminal jet sources and it had been suggested that
these features are related to the jet formation mechanism. This now
appears unlikely, and instead their presence may be
related to the viewing angle of the system, or they may be a common
feature of LMXRB systems as suggested by the recent detection of a
narrow ($\sigma < 70$~eV) absorption feature with an energy of
$7.01 \pm 0.03$~keV from 
GX\thinspace13+1 (Ueda et al.~\cite{u:01}). Here we report the discovery
of narrow X-ray absorption features from highly
ionized Fe in the XMM-Newton spectrum of \src.  

\begin{figure}
 \centerline{\includegraphics[width=6.3cm,angle=-90]{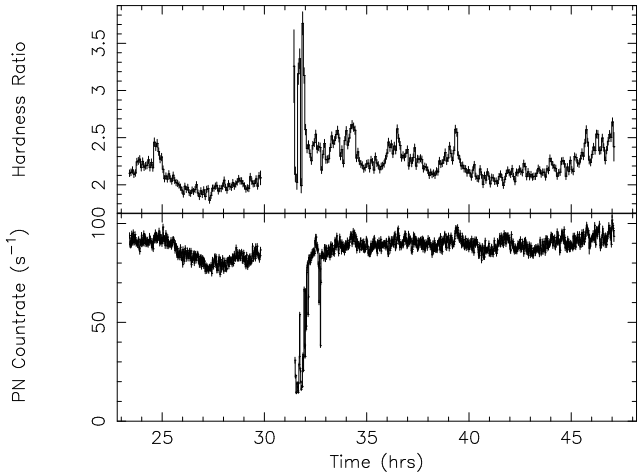}}
  \caption[]{The EPIC PN 2--10 keV lightcurve of \src\ with a binning
  of 64~s. The upper panel shows the hardness ratio
  (counts between 4--10~keV divided by those between 2--4~keV)
   with a binning of 256~s. Time
   is hours of 2001 February 11}
  \label{fig:lightcurve}
\end{figure}

\section{Observations}
\label{sect:obs}

The XMM-Newton Observatory (Jansen et al. \cite{j:01}) includes three
1500~cm$^2$ X-ray telescopes each with an EPIC at the focus. 
Two of the EPIC imaging spectrometers use MOS CCDs (Turner et al.
\cite{t:01}) and one uses a PN CCD (Str\"uder et al. \cite{st:01}).
The region of sky containing \src\ was observed by XMM-Newton
on 2001 February 11 23:23 UT to February 12 23:06~UT as part of the Performance
Verification Programme.
In order to minimize the effects of pile-up the PN and one MOS
were operated in 
their Small Window modes. The other MOS CCD was operated in its normal
Full Frame mode and the data are strongly affected by pile-up and
not discussed further. Due to the large amount of absorption
there are too few counts in the 0.35--2.5~keV RGS to perform a useful 
analysis. The OM was not operated during the observation. Since useful
data on \src\ was only obtained with one MOS, we concentrate on the
analysis of PN data, and use the lower count rate MOS data to 
check for consistency.

Raw data products were extracted from
the public XMM-Newton archive and then reprocessed using
version 5.2 of the Science Analysis Software (SAS),
before being further filtered using {\sc XMMSELECT}. For the MOS
only X-ray events corresponding to patterns 0--12 were selected.
Known hot, or flickering, pixels and electronic noise were rejected 
using the SAS. 
Source events were extracted from circular regions of 60\arcsec\
radius centered on \src.
Background spectra were obtained from a total of $\sim$60 ks
observations of several fields in the normal Full Frame modes. 
Background subtraction, however, is not critical for such a
bright source. 
All quoted count rates and fluxes are corrected for
instrumental deadtime. 

\begin{figure}
%  \hbox{\hspace{0.0cm}
   \includegraphics[width=5.5cm,angle=-90]{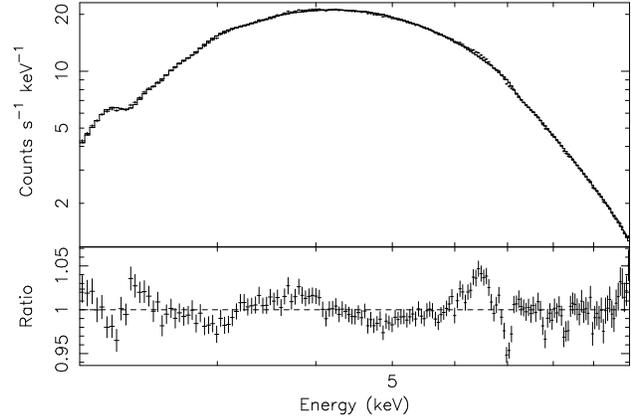}
  \caption[]{The PN spectrum of the \src\ persistent emission
             and the best-fit blackbody and power-law 
             continuum model together with a broad emission line.
             The lower panel shows the ratio of model to data counts
             when the normalizations of the line features are set to zero}
  \label{fig:spectrum}
\end{figure}

\begin{figure}
\hbox{\hspace{0.4cm}
\includegraphics[width=6.2cm,angle=-90]{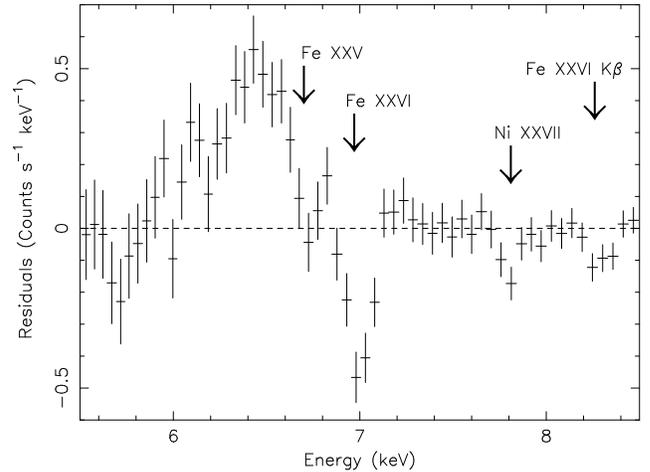}}
\caption[]{The 5.5--8.5~keV residuals from the fit shown in
Fig.~\ref{fig:spectrum}.
A broad emission feature at 6.58~keV
and 2 narrow features identified with Fe\,{\sc xxv} K$\alpha$ and 
Fe\,{\sc xxvi} K$\alpha$ 
absorption at 6.72 and 7.00~keV are clearly evident. Fainter
features at 7.83 and 8.28~keV may be present and 
are tentatively identified with Ni\,{\sc xxvii} K$\alpha$ and
Fe\,{\sc xxvi} K$\beta$ absorption}
\label{fig:line_residuals}
\end{figure}

\section{Results}
\subsection{X-ray lightcurve}

The 2--10~keV \src\ 
lightcurve obtained from the PN during the 2001 February
observation is shown in Fig.~\ref{fig:lightcurve} with a binning of
64~s. The upper panel 
shows the PN hardness ratio (counts in the energy range 
4--10~keV divided by those between 2--4~keV) with a binning of
256~s. A dip started after 2001 February 12 05:47~UT
and had a duration of $<$3~hours.
During the dip the source exhibited rapid, irregular changes in
count rate and hardness ratio. 
Such behavior
is typical of \src\ (see e.g., Ba\l uci\'nska-Church et al. \cite{bc:00};
Smale et al.~\cite{s:01}). 
During the persistent emission some flaring, as investigated in detail
by Smale et al.~(\cite{s:01}) and
Ba\l uci\'nska-Church et al.~(\cite{bc:01}) who show that it
is primarily a high-energy effect, not visible in RXTE Proportional Counter
Array 2--6~keV data, but becoming
dominant at energies $\approxgt$10~keV. 

\subsection{Persistent emission spectrum}
\label{subsect:spectrum}

The persistent emission spectrum of \src\ was first investigated by
excluding source events between 2001 February 12 05:47 and 09:06~UT
during the dip. This selection results in a
2.0--10~keV PN count rate of 89~s$^{-1}$ and an exposure of
52~ks. This count rate is close to the 100~s$^{-1}$ level, 
above which pile-up effects become important (Str\"uder et al. \cite{st:01}).
At high count rates the energies of multiple events are added and 
assigned to a single event. This results in harder spectral
continua than in the unaffected case, but should not produce additional
line features. 
In the PN Small Window mode only a 63 x 63 pixel region of the
central CCD is read out with a time resolution of 5.7~ms. 
In order to investigate the effects of pile-up a second
spectrum was accumulated with the region contained 
within the central 11\arcmin\ radius of the
point spread function (PSF) excluded. This results in the
loss of 45\% of the counts.
Both spectra were
rebinned to oversample the full width half-maximum of the energy
resolution by a factor 3 and to have additionally a minimum of 20
counts per bin to allow use of the $\chi^2$ statistic.
In order to account for systematic effects a 2\%
uncertainty was added quadratically to each spectral bin.
The photo-electric absorption cross sections of Morrison \& McCammon
(\cite{mc:83}) are used throughout. 
All spectral uncertainties are given at 90\%
confidence, unless indicated otherwise.

The overall continuum was modeled using absorbed blackbody and 
power-law components. 
A power-law was chosen rather
that a cutoff power-law or Comptonization model since the cutoff
energy of $\sim$12~keV (Ba\l uci\'nska-Church et al., \cite{bc:00})
is too high to be important in the EPIC energy range.
This combination models the overall shape of the continuum moderately 
well and gives $\chi ^2$s of 517.5 and 483.6 for 173 
degrees of freedom (dof) for the total and PSF excluded
spectra, respectively. 
A comparison of the best-fit parameters 
shows that, as expected, the total spectrum is significantly harder than
the PSF excluded one. We therefore use the PSF excluded spectrum
to determine the continuum parameters.

Two broad 
spectral features 
remain in both spectra centered on 2.5 and 3.5~keV.
The first feature is almost certainly
due to an incorrect modeling of the Au mirror edges,
while the second feature, with a peak-to-peak amplitude of 5\% is likely also
due to remaining PN calibration uncertainties and
is not visible in the MOS spectrum.
Examination of the remaining fit
residuals in both spectra
shows that broad positive residuals are present near
6.6~keV, as well as two narrow negative residuals at slightly higher energies.
The properties of these features are consistent between the
two spectral fits. Since pile-up is not expected to significantly affect
narrow spectral features, we use the results obtained from the fit
to the total spectrum to investigate the properties of any discrete
spectral features.
The addition of a $\sigma = 0.23 \pm 0.035$~keV 
broad emission line at $6.39 \, ^{+0.03} _{-0.04}$~keV with
an equivalent width (EW) of $33 \, ^{+6} _{-5}$~eV reduced the $\chi ^2$ to
299.7 for 170 dof. When modeled in this way, this feature is 
consistent with that seen by 
previous missions such as ASCA which measured an energy of
$6.4 \, ^{+0.1} _{-0.3}$~keV and an EW of $13 \, ^{+13} _{-9}$~eV 
(Asai~et al.~\cite{a:00}).
The F-statistic value of 41.2 indicates that
the probability of such a decrease occurring by chance is $5 \, 10^{-20}$.
The narrow features were modeled as Gaussian absorption lines.
Including the first feature at $7.00 \pm 0.02$~keV with an EW
of $-16.6 \, ^{+1.9} _{-5.9}$~eV reduced the $\chi ^2$ to 249.0 for 167 dof. 
This decrease in $\chi ^2$ 
corresponds to an F-statistic of 11.3. The probability of such
a decrease occurring by chance is $8 \,10^{-7}$. 
Including a second narrow feature at $6.72 \pm 0.03$~keV with an EW of 
$-7.5 \, ^{+1.7} _{-6.3}$~eV 
reduced the $\chi ^2$ to 218.9 for 164 dof. This decrease in
$\chi ^2$  corresponds to an F-statistic of 7.5. The probability of such
a decrease occurring by chance is $10^{-4}$.
When the two absorption lines are included in the fit, the properties of the 
emission
feature change to a best-fit energy of 
$6.58 \, ^{+0.07} _{-0.04}$~keV, a width ($\sigma$) of 
$470 \pm 70$~eV and an EW of $78 ^{+19} _{-6}$~eV.
Fig.~\ref{fig:spectrum} shows the best-fit spectral fit
model for the \src\ total spectrum. The best-fit
parameters derived from the fit to the PSF core excluded spectrum 
consist of a $1.18 \, ^{+0.01} _{-0.02}$~keV
blackbody with a radius of {\bf $9.60 \, ^{+0.04} _{-0.10}$}~km 
(for a distance of
15~kpc), together with a power-law with a photon index, $\alpha$, of
$2.02 \pm 0.12$. Both components are modified by absorption equivalent
to $(7.4 \, ^{+0.4} _{-0.2}) \, 10^{22}$~atom~cm$^{-2}$.  
The best-fit continuum and line model parameters are given in  
Table~\ref{tab:spectrum} and the residuals in the region around
the Fe features are shown in Fig.~\ref{fig:line_residuals}.

Fig.~\ref{fig:line_residuals} also shows evidence for the presence
of two additional narrow absorption features
at $7.83 \, ^{+0.03} _{-0.05}$ and 
$8.28 \, ^{+0.07} _{-0.04}$~keV  
with EWs of $-1.4 \, ^{+1.4} _{-1.5}$ and $-1.7 \, ^{+1.1} _{-2.2}$~eV,
respectively (at 68\% confidence).
The measured energies are consistent with those
of Ni\,{\sc xxvii} K$\alpha$ at 7.806~keV and Fe\,{\sc xxvi} K$\beta$
at 8.26~keV, both features
seen in the ASCA spectrum of GRS\,1915+106 by Kotani et al.~(\cite{k:00}).
The reductions in $\chi ^2$ when these features are added to the best-fit
spectral model are however not significant. 
We caution also that 
this is a spectral region where the EPIC calibration is still relatively
uncertain, and so we regard the presence, and proposed identifications,
of the features as tentative.
We have searched for absorption edges in the PN spectrum
due to neutral Fe, Fe\,{\sc xxv}, and Fe\,{\sc xxvi} expected
at 7.12, 8.83 and 9.28~keV.
The 95\% confidence upper-limit to the optical depth of an edge
at 7.12~keV is 0.009 and there is some evidence for the presence
of edges at 8.83 and 9.28~keV with optical depths of $0.023 \pm 0.012$
and $0.011 \, ^{+0.013} _{-0.011}$.

\begin{table}
\begin{center}
\caption[]{Best-fit to the 2.0--10~keV XMM-Newton PN persistent emission
spectrum of \src\ using blackbody and power-law continuum components,
together with a broad emission line and two narrow absorption lines.
The (absorption corrected) 2--10~keV 
luminosity, $L$, and blackbody radius assume a 
distance of 15~kpc (Christian \& Swank~\cite{cs:97}). 
The continuum parameters (including \nh) were determined from
the PSF core excluded spectrum. The power-law normalization
and blackbody radius were corrected for the
missing flux (see text)}
\begin{tabular}{llr}
\hline
\noalign {\smallskip}
Component & Parameter              &  Value \\
\noalign {\smallskip}
\hline
\noalign {\smallskip}
& $N{\rm _H}$  ($10^{22}$~atom cm$^{-2}$) & $7.4 \pm ^{0.4}_{0.2}$ \\

&$L$ (\ergsec)    &   $3.8 \, 10^{37}$  \\

%%&$\chi ^2$/dof            & 218.9/164 \\
&$\chi ^2$/dof            & 195.6/164 \\

Blackbody &$kT$ (keV)     & $1.18 \, ^{+0.01} _{-0.02} $ \\
          &Radius (km)    & $9.60 \, ^{+0.04} _{-0.10}$ \\

Power-law&$\alpha$      & $2.02  \pm 0.12$ \\
          &Normalization at 1 keV& $0.15 \pm ^{0.01}_{0.07}$ \\

Fe emission & $E_{{\rm line}}$ (keV)     & $6.58 \, ^{+0.07} _{-0.04}$\\
feature&$\sigma$ (eV)  & $470 \pm 70$ \\
&$I_{{\rm line}}$ ($10^{-4}$~ph cm$^{-2}$ s$^{-1}$)&  $11.7 \, ^{+2.9} _{-0.9}$ \\
&EW              (eV)           &  $78 \, ^{+19} _{-6}$  \\

Fe\,{\sc xxv} abs& $E_{{\rm line}}$ (keV)         & $6.72 \pm{0.03}$ \\
feature&$\sigma$ (eV)                  &   $<$50    \\
&$I_{{\rm line}}$ ($10^{-4}$~ph cm$^{-2}$ s$^{-1}$)
&  $-1.2 \, ^{+0.3} _{-1.0}$\\
&EW              (eV)           & $-7.5 \, ^{+1.7} _{-6.3}$ \\

Fe\,{\sc xxvi} abs& $E_{{\rm line}}$ (keV)         & $7.00 \pm{0.02}$ \\
feature&$\sigma$ (eV)                  &   $<$56    \\
&$I_{{\rm line}}$ ($10^{-4}$~ph cm$^{-2}$ s$^{-1}$)
&  $-2.3 \, ^{+0.3} _{-0.8}$\\
&EW              (eV)           & $-16.6 \, ^{+1.9} _{-5.9}$ \\

\noalign {\smallskip}                       
\hline
\label{tab:spectrum}
\end{tabular}
\end{center}
\end{table}

\begin{figure}
  \hbox{\hspace{0.0cm}
   \includegraphics[width=6.0cm,angle=-90]{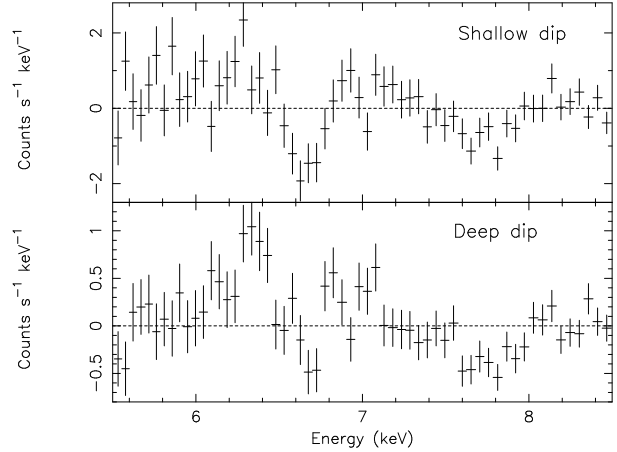}}
  \caption[]{Residuals in the 5.5--8.5~keV energy range
   when a power-law and blackbody
  continuum is fit to the shallow (upper panel) and deep (lower
  panel) dip spectra. The complex nature of the remaining structure is
  evident}
  \label{fig:dips_res}
\end{figure}

\subsection{Orbital dependence of the spectral features}
\label{sect:orbital}

In order to investigate whether the properties of the absorption features
depend on orbital phase, the persistent emission interval was divided
into three approximately equal duration parts, one before the dip, and
two after (see Fig.~\ref{fig:lightcurve}). The approximate phases covered
are 0.57--0.88, 0.03--0.37 and 0.37--0.70 using the ephemeris of 
Smale et al.~(\cite{s:01}), which has an uncertainty of $\sim$0.03 at the
time of the XMM-Newton observation (phase 0.0 corresponds to the dip
centers).
The same continuum model as used in Sect.~\ref{subsect:spectrum}
was fit to the 3 PN spectra and the
EWs of the two Fe absorption features determined. For the deeper 
7.00~keV feature, EWs of $-23.4 \, ^{+5.2} _{-7.0}$,  
$-24.1 \, ^{+5.3} _{-6.4}$, and $-15.7 \, ^{+4.6} _{-8.5}$~eV were
obtained. Thus, there does not appear to be any obvious dependence of the
EW of the Fe\,{\sc xxvi} feature on orbital phase, at least outside of
dipping intervals. A similar analysis, when applied to the shallower
Fe\,{\sc xxv} feature is not so constraining, but gives a similar result.

The properties of the absorption features during the dip were 
next investigating by accumulating PN spectra during intervals of
deep and shallow dipping activity. Events were selected by
making a lightcurve with 16~s resolution of the dipping interval. Time
intervals corresponding to PN count rates above and below 37~s$^{-1}$
were then derived. 
Deep and shallow dip spectra were accumulated using
these time intervals.
The 2--10~keV count rates are 23 and 61~s$^{-1}$ for exposures of
1047 and 480~s. 
In order to characterize the continua, the same blackbody and
power-law model as in Sect.~\ref{subsect:spectrum} with all 
spectral parameters allowed to vary was used to give
$\chi ^2$ values of 198.3 and 172.2 for 167 and 164 dof, for the
shallow and deep dip spectra, respectively. 
It is well known that detailed modeling of the spectral 
changes during dips requires multiple components
which undergo different amounts of absorption and covering (e.g., 
Church \& Ba\l uci\'nska-Church~\cite{c:95}). However, the simple
approach adopted here allows us to investigate the properties of
any narrow absorption features present.

Fig.~\ref{fig:dips_res} shows the residual in the energy range
5.5--8.5~KeV obtained in this way for the shallow and deep dipping
intervals. There is a great deal of spectral structure
remaining. The shallow dip spectrum
shows evidence for two deep features which may be modeled
as absorption lines with energies of $6.66 \pm 0.03$ and $7.71\pm0.06$~keV
(Table~\ref{tab:dips}).
The lower energy feature may be Fe\,{\sc xxv}~K$\alpha$ resonant
absorption, while the energy of the second feature does not correspond
to that of any likely line or edge feature. 
The same two features are also 
present in the deep dip spectrum, together with an
emission feature at $6.47 \, ^{+0.09} _{-0.15}$~keV. This may be the 
same Fe emission feature as seen in the persistent emission spectrum.
The deep Fe\,{\sc xxvi}~K$\alpha$ feature at 7.00~keV 
seen in the persistent emission spectrum may also be present in both 
dip spectra (see Fig.~\ref{fig:dips_res}),
but is much shallower than the feature
identified with Fe\,{\sc xxv}~K$\alpha$. This suggests that during
dipping the additional absorbing material is less strongly ionized
than that responsible for the absorption features seen in the persistent
emission. A detailed
study of such features may allow the ionization state, velocity and
temperatures of the material responsible for the dips to be probed.

\begin{table}
\begin{center}
\caption[]{Spectral parameters for the line emission
during dipping intervals. The continuum model used is a power-law 
and a blackbody. $E$, $\sigma$ and EW are the line energy, width and
equivalent width, respectively. A negative EW indicates an absorption
feature. An emission line is only required in the fit to the
deep dipping spectrum}
\begin{tabular}{lccc}
\hline
\noalign {\smallskip}
\mc{2}{l}{Shallow dipping} & &     \\
\hline
\noalign {\smallskip}
$E$ (keV)       & \dots & $6.63 \pm 0.03$ & $7.71 \pm 0.06$ \\
EW (eV)      & \dots & $\llap{$-$}66 \pm 18$    
& $\llap{$-$}151 \, ^{+44} _{-35}$ \\ 
$\sigma$ (eV)& \dots & $40 \pm 40$     & $180 \pm 70$ \\
\noalign {\smallskip}
\hline
\noalign {\smallskip}
\mc{2}{l}{Deep dipping}   &   &    \\
\hline
\noalign {\smallskip}
$E$ (keV)       & $6.47 \, ^{+0.09} _{-0.15}$ & $6.63 \pm 0.06$ & 
$7.74 \, ^{+0.06} _{-0.14}$ \\
EW (eV)      & $370 \pm 110$ & $\llap{$-$}78 \, ^{+43} _{-70}$ 
& $\llap{$-$}120 \, ^{+40} _{-150}$ \\
$\sigma$ (eV) & $370 \, ^{+390} _{-110}$ & $80 \pm 80$ & $140 \, ^{+150} _{-70}$ \\
\noalign {\smallskip}
\hline
\noalign {\smallskip}
\label{tab:dips}
\end{tabular}
\end{center}
\end{table}

\begin{figure}
\hbox{\hspace{0.0cm}
\includegraphics[width=7.5cm,angle=0]{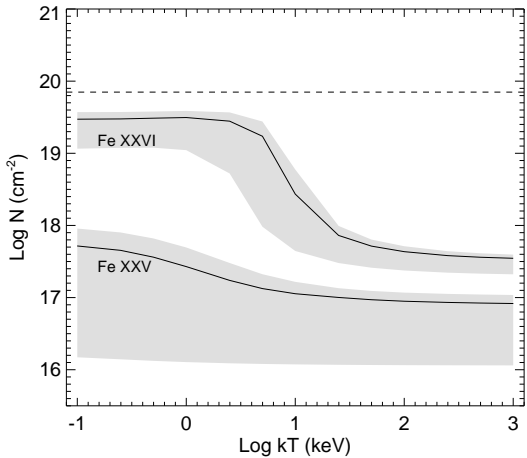}}
\caption[]{Fe column densities derived from the curve
of growth analysis for the Fe\,{\sc xxv} 
and Fe\,{\sc xxvi} absorption lines observed in the persistent
emission. 
The solid lines show the
values derived using the best-fit values of the EWs and the shaded areas
show the 90\% confidence regions.
The dotted line
indicates the Fe column density equivalent to one Thomson depth}
\label{fig:cofg}
\end{figure}

\section{Discussion}
\label{sect:discussion}

Narrow X-ray absorption lines were first detected from the superluminal
jet sources GRO\,J1655$-$40 (Ueda et al.~\cite{u:98}; 
Yamaoka et al.~\cite{y:01}) and GRS\,1915+105 (Kotani et al.~\cite{k:00};
Lee et al.~\cite{l:02}). ASCA observations of GRO\,J1655$-$40 revealed the
presence of absorption features due to Fe\,{\sc xxv} and Fe\,{\sc xxvi} which
did not show any obvious dependence of their EWs on orbital phase.
GRO\,J1655$-$40 has been observed to undergo deep absorption dips 
(Kuulkers et al.~\cite{k:98}) consistent with observing the source
at a high inclination angle of $60\degmark$--$75\degmark$ 
(e.g., Frank et al.~\cite{f:87}). ASCA observations of GRS\,1915+105
revealed, in addition, absorption features due to Ca\,{\sc xx},
Ni\,{\sc xxvii} and Ni\,{\sc xxviii}. A recent {\it Chandra} HETGS observation
of this source revealed absorption edges of Fe, Si, Mg, and S
as well as resonant absorption features from
Fe\,{\sc xxv} and Fe\,{\sc xxvi} and possibly Ca\,{\sc xx}
(Lee et al.~\cite{l:02}). 
Until recently, it was possible that these absorption features were
peculiar to superluminal jet sources and related in some way to the jet
formation mechanism. With the discovery of narrow absorption features from
the LMXRBs GX\thinspace13+1 (Ueda et al.~\cite{u:01}), 
MXB\thinspace1658$-$298 (Sidoli et al.~\cite{si:01}) and now
\src, this appears not to be the case, and as proposed
by Kotani et al.~(\cite{k:00}) ionized absorption features
may be common characteristics of systems accreting via a disk. 

In order to estimate the column densities of Fe that produced
the absorption lines observed in the persistent emission
we performed a curve of growth analysis
following that of Kotani et al.~(\cite{k:00}).
This allows the EWs of the lines to be converted into column densities
depending on the kinematic temperature of the absorbing material.
The kinematic temperature includes contributions from thermal 
motions as well as any bulk motions.
An Fe abundance of $4.7 \, 10^{-5}$ was assumed
(Anders \& Grevesse~\cite{a:89}).
Fig.~\ref{fig:cofg} shows the calculated columns for
the Fe\,{\sc xxv} and Fe\,{\sc xxvi} features
for a range of assumed temperatures. 
As expected, a lower kinematic temperature requires a higher ion
column density, and hence a higher hydrogen column density. 
With the range of assumed temperatures, the derived Fe column densities
are always less than the \nh\ corresponding 
to one Thomson optical depth ($1.5 \, 10^{24}$~atom~cm$^{-2}$). Such a high
column (shown as a horizontal line on Fig.~\ref{fig:cofg}) 
would result in the absorption lines being
strongly diminished. Examination of Fig.~\ref{fig:cofg}
shows that the total Fe column density must be 
$\approxgt$$10^{17.3}$~cm$^{-2}$ for the range of assumed temperatures.

The total line EW observed from \src\ is a factor $\sim$3 
lower than that observed
from the eclipsing, dipping source MXB\thinspace1658$-$298 by 
Sidoli et al.~(\cite{s:01}) also using XMM-Newton. In addition, the
ratio of H-like to He-like line EWs of $\sim$2.2 in \src\ is higher than in
MXB\thinspace1658$-$298 where a ratio of $\sim$1.3 was determined. 
This suggests that there is relatively more Fe\,{\sc xxvi} than Fe\,{\sc xxv} 
present in the absorber in \src, compared to MXB\thinspace1658$-$29.
There could be a number of reasons for these differences. (1)
The overall metal abundance could be higher in MXB\thinspace1658$-$298
than in \src, resulting in deeper absorption features. (2) The
geometries of the absorbing regions could be different, resulting
in different fractions of scattered photons entering the line of
sight. (3) As implied by the larger ratio of Fe\,{\sc xxvi} to Fe\,{\sc xxv} 
EW in \src\ compared to MXB\thinspace1658$-$298 the ionization parameter
of the absorbing system may be higher in this system and much of the Fe
in the absorber may be fully ionized. (4) Since MXB\thinspace1658$-$298 shows
dips {\it and} eclipses, while \src\ only shows dips, it is possible
that MXB\thinspace1658$-$298 is viewed closer to the orbital plane than
\src. This could mean that the line of sight may pass through more
material. It will be very interesting to obtain column
densities of absorption features in other LMXRB (dip-)sources to 
investigate which of the above possibilities is true.

Asai et al.~(\cite{a:00}) report on a spectral survey of Fe-K emission
lines from 20 LMXRB with ASCA. Significant lines are detected from about
half the sources. The mean line energy is 
$6.56\pm 0.01$~keV with a full width half-maximum 
of $\sim$0.5~keV. Neither the line energy, EW nor width appear
to depend on source luminosity or type. When the \src\ persistent
emission PN spectrum is fit with a single Gaussian emission feature, the 
best-fit energy of $6.39 \, ^{+0.03} _{-0.04}$~keV is significantly below
this mean value. However, when the two absorption features are included,
then the best-fit energy of $6.58 \, ^{+0.07} _{-0.04}$~keV is fully
consistent with the mean energy found by ASCA for the LMXRB sample. 
This difference well illustrates the dependence of derived spectral
results on the chosen model. It also suggests that other LMXRB with
Fe-line energies well below the average in the ASCA sample may exhibit
absorption features. Examination of Table~5 of Asai et al.~(\cite{a:00})
shows that the line observed from XB\,1916$-$053 has the lowest energy
in the sample of $5.9 \, ^{+0.2} _{-0.1}$~keV. Interestingly, XB\,1916$-$053 is
another dip source (White \& Swank~\cite{ws:82}), supporting the idea that
the presence of absorption features may depend on viewing angle.

\begin{acknowledgements}
Based on observations obtained with XMM-Newton, an ESA science mission
with instruments and contributions directly funded by ESA member states
and the USA (NASA).  
L. Boirin acknowledges an ESA Fellowship. We thank
T.~Kotani for making
his curve of growth software available.

\end{acknowledgements}

\end{document}